Size-independent differences between the mean of discrete stochastic systems and the corresponding continuous deterministic systems

Chetan J Gadgil
Chemical Engineering Division, National Chemical Laboratory,
Dr Homi Bhabha Road, Pune 411008, India

Email: cj.gadgil@ncl.res.in
Phone: +91-20-25902163   Fax: +91-20-25902612

## *Abstract*

In this paper I show that, for a class of reaction networks, the discrete stochastic nature of the reacting species and reactions results in qualitative and quantitative differences between the mean of exact stochastic simulations and the prediction of the corresponding deterministic system. The differences are independent of the number of molecules of each species in the system under consideration. These reaction networks are open systems of chemical reactions with no zero-order reaction rates systems. They are characterized by at least two stationary points, one of which is a nonzero stable point, and one unstable trivial solution (stability based on a linear stability analysis of the deterministic system). Starting from a nonzero initial condition, the deterministic system never reaches the zero stationary point due to its unstable nature. In contrast, the result presented here proves that this zero-state is the only stable stationary state for the discrete stochastic system. This result generalizes previous theoretical studies and simulations of specific systems and provides a theoretical basis for analyzing a class of systems that exhibit such inconsistent behavior. This result has implications in the simulation of infection, apoptosis, and population kinetics, as it can be shown that for certain models the stochastic simulations will always yield different predictions for the mean behavior than the deterministic simulations.

## *Keywords*

Stochastic reaction kinetics; deterministic chemical kinetics; biochemical reaction kinetics; discrete systems; continuous systems

## Introduction

There are two prominent approaches to the theoretical study of the dynamics of biological systems: the continuous deterministic kinetics (CDK) approach and the discrete stochastic kinetics (DSK) approach. Both these approaches treat time as a continuous variable. In the former, the rate of change of species concentration is expressed as a system of ordinary differential equations, with the concentrations treated as continuous variables. Integration of these equations yields the concentration of each species as a function of time. In the latter (DSK), an equation (called the master equation) is derived for the probability that the system is in a particular state (defined by the number of molecules of every species) at a given time. This equation includes terms for the rate of change of this probability due to reactions that take the system away from this state, and reactions that give rise to this state. There are methods that are in between these approaches, e.g. continuous stochastic kinetics wherein the dynamics are modeled using stochastic differential equations, and the Boolean dynamics approach which is a discrete deterministic approach. CDK formulations have been previously referred to as chemical kinetics equations (Goutsias, 2007) or classical chemical kinetics (Samoilov and Arkin, 2006); and DSK approaches have been referred to as chemical master equation (Goutsias, 2007; Samoilov and Arkin, 2006). Since the objective of the paper is to explain differences that result from the discrete and stochastic (as opposed to continuous and deterministic) nature of the system, I will use the DSK/CDK terminology.

I will now briefly describe the equations for the CDK and DSK approaches. Further description of this formulation and its extension in terms of reaction extents has been described earlier (Gadgil et al., 2005). Consider the set of $N_R$ reactions between the $N_S$ species $M_1, M_2, ..., M_{N_S}$, given by

$$\sum_i^{reac} -v_{ij}^{reac} M_i = \sum_i^{prod} v_{ij}^{prod} M_i \text{ for the } N_R \text{ reactions j=1,2,...,}N_R \qquad (1)$$

The coefficients $v_{ij}$ represent the stoichiometric coefficients of the i$^{th}$ species in the j$^{th}$ reaction (negative for reactants). In this formulation, forward and reverse steps of a reversible reaction are treated as two individual reactions. The rate of change of species concentrations $c \equiv \left[ c_1, c_2, ..., c_{N_S} \right]^T$ for the $N_S$ species is given by

$$\frac{dc}{dt} = \nu \Re(c) \tag{2}$$

where $\nu$ is the ($N_R$ x $N_S$) stoichiometric matrix composed of the stoichiometric coefficients $\nu_{ij}$ and $\Re(c)$ is the non-negative vector (of length $N_R$) of reaction rate expressions. For mass action kinetics, the rate expression for the j$^{th}$ reaction is given by

$$\Re_j(c) = k_j \prod_i^{reac} c_i^{-\nu_{ij}} \tag{3}$$

In terms of the number of molecules of each species $n_i$, equation (2) may be written as

$$\frac{d(n_i/N_A V)}{dt} = \nu \; \Re(n_i/N_A V) \tag{4}$$

where $N_A$ is Avogadro's number, and $V$ is the system volume. This differential equation representation of the dynamics of the reaction system represented by (1) is the CDK formulation, and may be integrated to obtain the evolution of the number of molecules of each species as a function of time and system constants (volume, reaction rate constants, initial conditions).

In the DSK approach, the number of molecules of each species at a given times is given by a probability P(n;t). The change in this probability is due to reactions from progenitor states that feed into this state, and reactions that take the system away from this state. This master equation describing the change in P is given by

$$\frac{dP(\mathrm{n};t)}{dt} = \sum_{\mathrm{m}_1 \in S(\mathrm{n})} \Re(\mathrm{m}_1 \to \mathrm{n}) P(\mathrm{m}_1;t) - \sum_{\mathrm{m}_2 \in \Omega(\mathrm{n})} \Re(\mathrm{n} \to \mathrm{m}_2) P(\mathrm{n};t) \tag{5}$$

where $S(\mathrm{n})$ and $\Omega(\mathrm{n})$ are the sets of progenitor and successor states of the state $\mathrm{n}$, i.e., their members are the states that are one reaction away from the state characterized by $\mathrm{n} \equiv (n_1, n_2, ... n_{N_S})$ molecules of species $M_1, M_2, ..., M_{N_S}$ respectively. Thus the occurrence of one (specific) reaction when the system is in any of the states present in $S(\mathrm{n})$ will take it to state $\mathrm{n}$, and the occurrence of any one reaction while in state $\mathrm{n}$ will take it to a state belonging to $\Omega(\mathrm{n})$. $\Re(m_1 \to \mathrm{n})$ is the rate expression for the progenitor reaction and is equivalent to $\Re(m_1)$ as used in (4). Numerically, the DSK approach considers each reaction given in (1) and, from the number of molecules of each species and the kinetic expression given by $\Re(n_i/N_A V)$, calculates the propensity of each reaction. From the sum of all propensities, the distribution function for the expected time for the next reaction is calculated. Random numbers are used to determine when the next reaction will occur and which reaction will occur at that time. Several instances of such numerical runs using independent sets of random numbers are used to generate a numerical probability distribution for the system. As the available computational power in terms of number of processors rapidly increases (Butler, 2007), such 'embarrassingly parallel' computations will become increasingly more feasible for large reaction systems.

There has been great interest in comparing the mean behavior of the DSK system to the result of the CDK formulation. Forger and Peskin (Forger and Peskin, 2005) simulated the circadian cycle using deterministic and stochastic simulations, and showed that the DSK dynamics lead to oscillations that are inaccessible to the CDK solution. Zheng and Ross (Zheng and Ross, 1991) investigated the evolution of the chemical intermediate in the autocatalytic Schlogl model using CDK and DSK approaches and found that, for a range of parameter values, the mean concentration calculated using DSK was always different than the CDK predictions. Srivastava and coworkers (Srivastava et al., 2002) investigated virus infection kinetics through a stochastic model, observed that the predicted mean of the DSK simulations differs from the prediction of the CDK simulation. This difference is attributed to the fact that individual stochastic simulations could access and remain at the CDK-unstable zero-state. Goutsias (Goutsias, 2007) has given a thorough review of previous efforts to compare CDK and DSK methods, and presented an approach that separates the dynamics into "macroscopic" and "mesoscopic" terms. This separation is akin to the concept of dividing system behavior into "average rate" and "stochastic rate" adopted by Gomez-Uribe and Verghese (Gomez-Uribe and Verghese, 2007). Goutsias has proved that when the mesoscopic term is zero, the mean of the DSK approach corresponds exactly to the CDK approach. A method of approximation of the mesoscopic term has also been presented. However, for nonlinear systems, an exact analytical expression for the mesoscopic term is not possible. Hence numerical computations are necessary to evaluate the extent of the difference in the predictions of the two approaches. Samoilov and Arkin (Samoilov and Arkin, 2006) systematically classify "deviant effects" occurring through various causes. They present an example of a system ($X+X \rightarrow Y+Y$; $X+Y \rightarrow X+X$) that has a zero steady state which is unstable according to the CDK formulation but stable according to the DSK formulation. They analyze the master equation for this specific system to prove that the $X=0$ state is the only steady state reached by the DSK formulation. They present an intuitive explanation of the phenomenon (if X runs out, no more X can be produced); and state that similar systems are expected to show this deviation between the DSK and CDK predictions, irrespective of the size of the system. However a formal proof for the result is not presented. Vellela and Qian (Vellela and Qian, 2007) consider the reaction A+X ↔ 2X; X → C for a case where the amount of A is held constant. Keizer had previously shown that for this system the only steady state predicted by the DSK formulation is the zero-state. Vellela and Qian term this phenomenon "Keizer's paradox" and present a detailed analysis of this reaction. They derive an expression for the time to extinction as a function of the initial number of molecules of X, and get the interesting result that the time quickly increases to a very high level after the number of molecules exceeds the steady-state number predicted by the CDK formulation. By setting the death rate when X=1 to zero, they compute the 'quasi-stationary' distribution that is reached at finite times. This thorough analysis presents a comprehensive comparison of the predictions of the DSK and CDK approaches as a function of time. They also reformulate the equations so that it is equivalent to the logistic growth model in population biology and analyze their results in the context of previous work on stochastic logistic growth models.

In the field of population modeling, it has been shown for a few specific models that the results of the stochastic analysis predict different behavior than the deterministic analysis. McKane and Newman (McKane and Newman, 2004) have analyzed individual based models (i.e. discrete stochastic models) of population dynamics and shown through simulations that the mean of DSK simulations is always less than the predictions of the corresponding CDK system, and the DSK system reaches extinction states that the CDK system cannot predict. Nicolis and Prigogine considered the Lotka-Volterra model corresponding to the reactions (X → 2X; X + Y → 2Y; Y → death) for the predator species X and prey species Y. They proved that a DSK analysis predicts that fluctuations lead the system away from the steady state predicted by the CDK model (Nicolis and Prigogine, 1971). Reddy (Reddy, 1975) proved that for this system, the only possible steady state predicted by the CDK model is the trivial one where predator and prey concentrations are zero. Other single-species population models have been analyzed (see for example Nassel's papers (Nasell, 1999; Nasell, 2001) for an analysis of the Verhulst and other logistic models) and methods for estimating the time to extinction have been presented. However this result has not been extended to general population dynamics models.

From these reports, it is clear that a distinction has to be made between stability as predicted by CDK (linear stability) analysis, and stability as predicted by the DSK approach. I refer to the stability of stationary points as CDK-stable and DSK-stable to reflect this distinction. In this paper, I consider a class of dynamical systems corresponding to chemical reaction networks that have the following characteristics:

(i) They are fully connected open systems with some species being produced from and/or degraded into a constant precursor pool.

(ii) The production of species is regulated such that $\Re(0) = 0$.

(iii) They are characterized by one or more stationary states, one of which is the zero-state.

The fully-connected nature specified in condition (i) stipulates that all species can access the precursor pools either directly or through intermediate species, i.e. a species Y may be formed from a species X that is formed from a precursor pool; and Y may be converted to a species Z that is then degraded. Condition (ii) implies that the rate of production of species from a pool of precursors is governed by at least one of the other species in the network so that there is no constant-rate (zero-order) or constitutive production. It also stipulates that no reaction is zero-order, i.e. does not depend on the concentration of any species in the network. Thus systems of equations corresponding to fed-batch or continuous reactors, where the feed rate is independent of the reactor concentrations, are excluded from this analysis. Systems with constitutive expression rate terms are also excluded. Certain closed systems may also be reformulated to be included in this analysis, and will be discussed later.

I prove that for this class of systems with properties (i-iii), *irrespective of system size*, the DSK approach will always lead to the prediction that the zero-state is the only stable steady state. All exact simulations will reflect this fact by converging to the zero-state provided the simulation is carried out till the mathematical steady state is reached. This result is independent of the CDK-stability of the zero-state, and hence systems where the zero-state is CDK-unstable will have clear qualitative and quantitative difference between the predictions of the CDK and DSK formulations. This result provides an framework for analysis of several previous reports of specific DSK systems which exhibit behavior that is analytically or numerically shown to be not "consistent in the mean" with CDK simulation results. This result will lead to a guideline for evaluating the performance of stochastic simulation algorithms, where consistency with the mean

CDK behavior has been previously assumed to be a criterion for the accuracy of simulation algorithms.

## *Results*

I prove that for systems of any size fulfilling conditions (i-iii) the only stable steady state for the DSK formulation is the zero-state. This result is then extended for certain closed systems, which may be reformulated so as to conform to conditions (i-iii).

### Open systems with a zero-state stationary point

*Theorem*:

A zero-state stationary point of a discrete stochastic dynamical system satisfying conditions (i-iii) representing an open chemical reaction network with no zero-order reactions is its only DSK-stable steady state.

*Proof:*

Consider the DSK formulation of a system represented by the probability balance equation (5), reproduced below:

$$\frac{dP(n;t)}{dt} = \sum_{m_1 \in S(n)} \Re(m_1 \to n) P(m_1;t) - \sum_{m_2 \in \Omega(n)} \Re(n \to m_2) P(n;t)$$

At steady state

$$\sum_{m_1 \in S(n)} \Re(m_1 \to n) P_{ss}(m_1;t) = \sum_{m_2 \in \Omega(n)} \Re(n \to m_2) P_{ss}(n;t) \qquad (6)$$

The subscript *ss* refers to the steady state probability distribution. From condition (ii), one stationary point of the system is the zero-state ($n = 0$). Consider this zero-state stationary point. Equation (6) written for this point is written as

$$\sum_{m_1 \in S(0)} \Re(m_1 \to 0) P_{ss}(m_1;t) = \sum_{m_2 \in \Omega(0)} \Re(0 \to m_2) P_{ss}(0;t) \qquad (7)$$

Each term $\Re(0 \to m)$ in the right hand side of (7) for every state $m$ consists of reaction rates for reactions leading to the formation of the various species. From condition (ii), all these terms can be set to zero as (a) the rate depends on at least one of the constituent species and, (b) for transitions from the zero-state, $\Re(0) = 0$. Thus the sum on the right hand side has to equal zero. Now, let us consider the left hand side of (7). All the states $m_1$ in the term $\Re(m_1 \to 0)$ always correspond to sets of molecule numbers that are nonnegative numbers with at least one positive number. Hence the rates $\Re(m_1 \to 0)$ are non-negative and in particular at least one of the rates is positive. This follows from condition (i) which specifies that at least one of the species is degraded into the common pool. Hence the only possible solution for the left hand side to be zero is

$$P_{ss}(m_1;t) = 0 \ \forall \ m_1 \in S(n=0) \tag{8}$$

Thus, at steady state the probability that the system is in any of the states that feed into the zero-state $(n=0)$ is zero. Now, consider one of these states $m' \in S(n=0)$. At steady state, from (6), a balance for the state $m'$ gives

$$\sum_{l_1 \in S(m')} \Re(l_1 \to m') P_{ss}(l_1;t) = \sum_{l_2 \in \Omega(m')} \Re(m' \to l_2) P_{ss}(m';t) \tag{9}$$

Here, $l_1$ and $l_2$ are any of the states in the set of progenitor ($S(m')$) and successor ($\Omega(m')$) states of $m'$, respectively. Now, from (8), $P_{ss}(m';t) = 0$. Hence all the terms in the right hand side of (9) are identically zero. From an analysis similar to that leading to equation (8), it is seen that $\Re(l_1 \to m') \geq 0 \ \forall l_1 \in S(m')$ and at least one of the rates is positive; and hence $P_{ss}(l_1;t) = 0 \ \forall \ l_1 \in S(m')$. Thus the steady state probabilities of all the states that lead into $m'$ are identically zero. This is true for all states $m'$ that are the progenitor states ($S(0)$) of the zero-state. This analysis can be carried out further for all states $k_1$ that are the progenitor states of any of the states $l_1$, and so on till we cover all the states that *directly or indirectly* are the progenitor states for the zero-state. The probability of all such states is zero. Now, by a theorem proved by Erdi and Toth, p64 (Erdi and Toth, 1989), the rate of decrease of concentration for any variable in a system of equations corresponding to a set of chemical reactions has to lack "negative cross-effects". Thus the rate of decrease of every species in the reaction network has to be a function of its own concentration. From this fact and condition (i), the zero-state has to be accessible at every point through a set of reactions possibly through intermediate species. Hence all the non-zero states of the system directly and indirectly feed into the zero-state, and we have proved that their steady state probabilities are zero. The only remaining state with a non-zero steady state probability is the zero-state itself, which therefore has a steady state probability of one. Thus it is the only DSK-stable steady state. This result is independent of the initial state, i.e.

the initial number of molecules of any species in the reaction, and hence independent of the system size.

*Application to previous models*

Having analytically proved these results for open systems with zero-states that are stationary points, I will discuss their application to one representative stochastic model. I consider the model of virus infection dynamics proposed by Srivastava et al (Srivastava et al., 2002). This model comprises of three differential equations for the template, genome and structural protein which satisfy the conditions (i-iii) described previously. The authors have indeed computationally discovered that the mean of the DSK simulation is consistently lower than the result of the corresponding CDK simulation due to a certain number of stochastic simulation instances ('aborted simulations') reaching the zero-state. They have further analyzed the effect of initial template concentration and simulation time on the number of instances reaching the zero-state and presented results showing that this fraction of aborted simulations increases with a decrease in the infection level (i.e. increases as the initial condition is in closer proximity to the zero-state) as well as the time for which the simulation is carried out. From the theory presented here, it is clear that at $t \to \infty$, all simulation runs will reach the zero-state, which is the only DSK-stable state. However, for computationally and biologically practical values of the simulation time, an increase in the fraction of instances reaching the zero-state is observed, consistent with the theoretical predictions.

In the light of these results, some of the assumptions implicit in CDK descriptions of chemical kinetics and biochemical processes have to be re-examined. As an example, consider the exponential-growth model for bacterial growth in the absence of substrate limitation. The cell concentration is represented by an equation of the form $d[X]/dt=k[X]$ where k is a positive constant. Even if first-order cell death is considered at a specific rate $k' < k$, the deterministic model $d[X]/dt=k[X]-k'[X]$ predicts exponential growth. In contrast, as this system fulfills conditions (i-iii), a DSK approach will predict that extinction is the only steady state solution. The approach of Vellela and Qian (Vellela and Qian, 2007) can be used to estimate the time to reach this extinction state from a given initial number of cells. This time increases with an increase in the initial number of cells. Such an assumption is implicit in Lotka-Volterra models, where the death rate of the prey species in the absence of predator is clubbed with the birth rate to form a net birth rate. In such models, if the birth and death processes for each species are explicitly included, the DSK prediction will be that extinction of both species is the only stable steady state.

## Closed systems reducible to those having zero-states

For closed systems, a zero-state is impossible to attain, as the total mass is conserved. Hence the number of molecules of at least one species is non-zero, as the total mass at steady state is equal to the total starting mass. However, this very fact allows the possibility of reducing closed systems such that they have the properties (i-iii). Consider the system analyzed by Samoilov and Arkin (Samoilov and Arkin, 2006), viz, (X+X→Y+Y; X+Y→X+X). Here the sum of number of molecules of X and Y at any time is constant, equal to their total number at the start of the reaction. As such, $n_Y=n_0-n_X$; and in order to analyze system dynamics it is sufficient to consider just the evolution of $n_X$. X is formed as a function of its own concentration and that of Y, and is degraded to Y as a function of its own concentration. It therefore satisfies the constraints of a source and sink term that is dependent on species concentrations, and $n_X=0$ is indeed a stationary point of the system. All possible states ($[n_X,n_Y]=[0,n_0],[1,n_0-1],[2,n_0-2],…,[n_0,0]$) are direct or indirect progenitors of the zero-state (in this case defined as $n_X=0$); and hence the system satisfies conditions i-iii. As such it will also be governed by the results of the theorem, i.e. the only DSK-stable state is ($[n_X,n_Y]=[0,n_0]$) as proved by Samoilov and Arkin. Using this theorem, however, it is seen that *any* closed system with characteristics similar to Samoilov and Arkin's system will have the zero-state as the only DSK-stable steady state. Formally, the required characteristics of the system are that conditions i-iii for open systems should hold for *any one* combination of the independent species numbers. Thus the Samoilov-Arkin system dynamics can be equally well described if the number of molecules of Y are tracked, but then the system will not have a zero-state that is a stationary point (since if $n_Y=0$ the net rate of change of Y is non-zero as Y is being formed at a finite rate and consumed at a zero-rate), and so will not satisfy condition (iii).

## *Discussion*

There have been several reports over the past five decades that analyze the differences between the predictions of models describing specific chemical reaction systems using discrete stochastic kinetics (DSK) and the corresponding continuous deterministic kinetics (CDK) approaches. To my knowledge, this is the first report that explains the size-independent differences in the DSK and CDK behavior of a class of chemical reaction models. In many of the previous analyses, the conclusion has been that the predictions of the DSK and CDK models will coincide when the number of reacting molecules is large (Erdi and Toth, 1989; Leonard and Reichl, 1990; McQuarrie et al., 1964; Thakur et al., 1978; Turner et al., 2004). In contrast, there have been a few reports that analytically (Reddy, 1975; Samoilov and Arkin, 2006; Vellela and Qian, 2007) or numerically (McKane and Newman, 2004; Srivastava et al., 2002) demonstrate results for specific systems where these predictions differ. In this report, I show that for a class of reaction systems (open systems with a zero-state stationary solution and regulated production rates), irrespective of the number of molecules in the system, the zero-state is the only DSK-stable state, and this fact is the cause of the observed differences in the predictions of DSK and CDK models.

Previous results on the "law of large numbers" have proved that the mean of the DSK calculations tend to the result of the CDK calculation when the number of reacting molecules is large, but this is true only in the limit as both number of molecules and volume tend to infinity, such that the concentration remains constant (Kurtz, 1972). For such systems with the number of molecules tending to infinity, the time required to reach the zero-state will also tend to infinity irrespective of the transition probabilities. Hence this result is not applicable for such systems, since the zero state even if mathematically a stationary point, will take infinite reaction time to be accessible to the system.

The time required to reach this DSK-stable zero-state will depend on the initial state of the system. Systems with more molecules at the initial state (i.e. where the initial state is distant from the zero-state) will require more transitions to take the system to the zero-state. An analogy may be drawn between this process and random walk on a lattice, where one of the points of the lattice is a sink and at least one path connects it to all the points. Given a long enough time (at $t \to \infty$) the random walker will always access every point on the lattice. However the time required to reach a particular point (say the sink point corresponding to the zero-state) will depend on the distance of the sink from the starting point and the transition probabilities along the path.

As is seen in the viral dynamics model (Srivastava et al., 2002), the mean of the DSK simulations will always be lower than the corresponding CDK predictions. This is a result of some DSK simulation runs reaching the stable zero-state which is inaccessible to the CDK simulation trajectory if it is a CDK-unstable steady state. If these runs are omitted from the analysis, the mean of the remaining runs will be closer to the CDK predictions, but will be not represent the true DSK dynamics of the system. Increasing the proximity of the initial conditions to the zero-state, or increasing the simulation time will result in a greater fraction of runs reaching the zero-state. An unsolved problem is the assessment of time-to-extinction for the general systems presented in this paper, which would be a significant extension of the Valella-Qian analysis (Vellela and Qian, 2007) for one-species logistic growth models.

I have assumed that the reaction rates are (1) not zero-order, and (2) independent of time. These assumptions are valid for reactions whose rates and transition probabilities are described by mass action kinetics. For instance, systems with a stationary zero-state where all species undergo first order degradation reactions will always converge to the zero-state for the CDK formulation. Systems that do not involve a constant (zero-order) rate of production of any component, or constant inflow rates of any component typically do not have any zero-order terms in their CDK models. This analysis is however not applicable for reactions occurring in fed-batch or continuous stirred tank reactors as the inflow process may be regarded as a zero-order production reaction for the species present in the input steam.

The accuracy of approaches that approximate exact simulations of the master equation for a reaction system is often evaluated by calculating the degree to which the results match the results obtained using Gillespie's exact method, or the mean behavior as calculated using the CDK approach (Haseltine and Rawlings, 2002; Rao and Arkin, 2003; Turner et al., 2004). In the light of this result, the chemical reaction systems used to carry out such evaluations should be carefully chosen such that their DSK behavior is not guaranteed to be different from the CDK behavior. In an effort to speed up the simulation of the stochastic behavior of chemical reaction systems, stochastic differential equations are used instead of DSK methods. Such systems, being continuous in nature, may not lead to the CDK-stable zero-state at the same frequency as the correct CDK approaches. For instance, in the case of a Lotka-Volterra model where the transition parameters are considered to be the sum of a mean positive value and a noise term, it has been shown (Mao et al., 2002) that the mean value is guaranteed to stay in the positive orthant, which contradicts the result of the DSK approach that predicts extinction. Hence stochastic differential equations should not be used while investigating the stochastic dynamics of such systems.

There are other fields where differences in CDK and DSK dynamics have been reported. These examples include several reports of dynamics of spatially distributed systems. The presence of multiplicative noise in a spatially distributed system has been shown to result in the system undergoing a transition to a symmetry breaking state which cannot be reached in the absence of the noise term (Van den Broeck et al., 1994). For spatially coupled system represented by logistic dynamics ($x_{n+1}^i = 1 - a(n,i)\, x_n^i$), it has been shown (Sinha, 1992) that the law of large numbers is not true when the environment ($a(n,i)$) varies randomly in time but is correlated in space, but does hold when it is either (a) constant in both time and space, (b) constant only in time, or (c) random in time and uncorrelated in space. In fact it has been suggested that care must be exercised while applying the law of large numbers to such systems (Ding and Wille, 1993). Systems exhibit different behavior depending on the nature of their interconnections. In a network of coupled chaotic oscillators modeling neuronal activity, as the coupling gets weaker, the network becomes spatiotemporally chaotic when the coupling connections are regular, but spatial synchronization is observed in the presence of random links (Jampa et al., 2007). In sociological network studies of the prisoner's dilemma type, for both direct ("I'll scratch your back and you scratch mine") and indirect ("I'll scratch your back and you scratch someone else's") reciprocity games, it has been shown that the dynamics are different when errors are introduced in implementing the rules (tit-for-tat, co-operation, defection) governing the behavior of the players. The presence of stochasticity results in different steady state populations of players (Brandt and Sigmund, 2006; Imhof et al., 2005). In models relevant to ecology, discreteness has been shown to be responsible for qualitative and quantitative changes in the interface dynamics of a model representing infection dynamics or bacterial growth. It has been suggested that a methodology that can account for the relevance of discreteness is a "necessary starting point" for the analysis of such systems (Kessler and Levine, 1998). I believe that the analysis presented in this paper can become the basis for such analysis of discrete reaction dynamics, and a logical extension would be to expand it for the analysis of distributed dynamical systems.

The final issue is the determination of which approach (CDK or DSK) to use: i.e. which is the biologically realistic model? It has been suggested (Wolkenhauer et al., 2004) that the rationale for choosing CDK or DSK should not be the numerical accuracy of the method that is employed, but whether the correct biological principal is reflected in the model. Based on these results, it seems that although the DSK approach leads to startling conclusions regarding the infinite-time behavior of some systems, it may be the better approach to simulate short (i.e. realistic) time dynamics. For example in the case of the virus dynamics model, it is conceivable that the 'aborted simulations' that reach the DSK-stable zero-state represent cells with spontaneous remission of the viral infection. Such zero-states (virus removal, apoptosis, extinction) may in fact represent biologically feasible states of the system that cannot be predicted by the CDK model. Thus, as stated previously (Nasell, 2001; Qian et al., 2002), the discrete stochastic kinetics model may represent a more complete kinetic description of the chemical kinetics, with continuous deterministic kinetics an approximation of the discrete stochastic kinetics.

## *Acknowledgements*

I thank Dr B. D. Kulkarni for his careful reading of the manuscript and for suggesting other areas where stochasticity changes system dynamics. I acknowledge financial support from National Chemical Laboratory, Pune through a start-up grant MLP 010926.